\documentclass[a4paper,11pt]{article}
\usepackage{pos}
\usepackage{bm}\allowdisplaybreaks[1]
\usepackage{graphicx, xcolor}
\usepackage{adjustbox}
\usepackage{tikz}
\usetikzlibrary{arrows.meta}
\usepackage{orcidlink}

\renewcommand{\sec}[1]{Sec.~\ref{#1}}
\newcommand{\fig}[1]{Fig.~\ref{#1}}
\newcommand{\eq}[1]{Eq.~\eqref{#1}}

\newcommand{\pp}[1]{\left(#1\right)}
\newcommand{\bb}[1]{\left[#1\right]}

\newcommand{\refcite}[1]{Ref.~\cite{#1}}
\newcommand{\refs}[1]{Refs.~\cite{#1}}

\newcommand{\beq}[1][]{\begin{equation}\label{#1}}
\newcommand{\eeq}{\end{equation}}
\newcommand{\bse}[1][]{\begin{subequations}\label{#1}}
\newcommand{\ese}{\end{subequations}}
\newcommand{\nn}{\nonumber}

\newcommand{\sgn}[1]{{\rm sgn}\left[#1\right]}
\newcommand{\wt}[1]{\widetilde{#1}}

\newcommand{\M}{\mathcal{M}}

\renewcommand{\H}{\mathcal{H}}
\newcommand{\Ht}{\widetilde{\mathcal{H}}}
\newcommand{\E}{\mathcal{E}}
\newcommand{\Et}{\widetilde{\mathcal{E}}}

\title{Overview on the theory and phenomenology of generalized parton distributions}

\author*[a,b]{Jian-Wei Qiu\,\orcidlink{0000-0002-7306-3307}\,}
\author[a]{Zhite Yu\,\orcidlink{0000-0003-1503-5364}\,}

\affiliation[a]{Theory Center, Jefferson Lab, Newport News, Virginia 23606, USA}

\affiliation[b]{Department of Physics, William \& Mary, Williamsburg, Virginia 23187, USA}

\emailAdd{jqiu@jlab.org}
\emailAdd{yuzhite@jlab.org}

\abstract{We give a brief overview on the theory and phenomenology of generalized parton distributions (GPDs), including the recently developed framework of single-diffractive hard exclusive process for matching GPDs to experimental observables. We concentrate on the extraction of GPDs from experimental processes, especially on the challenges and potential solutions regarding the separation of different GPDs and the extraction of $x$-dependence of GPDs, which is critically important for constructing the tomographic images and matching the $x$-moments of GPDs to various emergent hadron properties.}

\FullConference{7th International Workshop on ``Transverse phenomena in hard processes and the transverse structure of the proton'' (Transversity 2024)\\
 3--7 June 2024\\
Trieste, Italy\\}

\begin{document}

\vbox {
	\hfill{\bf JLAB-THY-24-4223}	
}
\vspace*{-2.5em}

\maketitle

\section{Introduction}
\label{sec:intro}
The research to explore internal structure of hadrons is entering a tomographic era 
owing to high-energy and high-luminosity accelerators together with theoretical and experimental advances to study
physical observables that have two distinct scales of momentum transfer in high energy collisions. 
With one hard scale $Q$ to localize the interaction with quarks and gluons (collectively referred to as partons) 
inside a hadron, a controllable soft scale $Q_s \ll Q$ allows to probe the motion and/or spatial distributions of 
quarks/gluons inside the hadrons.  Depending on if the probed hadron is broken or not during the hard collisions, 
there are two distinct classes of two-scale observables:  {\it inclusive} with the probed hadron broken and {\it exclusive} 
with the hadron kept intact.  The two-scale {\it inclusive} observables
provide opportunities to probe transverse-momentum-dependent parton distributions (TMDs), $f(x,k_T)$, 
with the active parton's longitudinal momentum fraction $x$ and transverse momentum $k_T\sim {\cal O}(Q_s)$~\cite{Boussarie:2023izj}. 
However, breaking the probed hadron necessarily triggers gluon radiation to introduce a challenge to match 
the active parton's $k_T$ detected in the collision to the confined motion of the same parton in 
the probed hadron before it was broken. 

On the other hand, the probed hadron of an {\it exclusive} process stays intact and is only diffracted from its momentum $p$
to $p'$ with the momentum transfer square $t=(p-p')^2$ that defines the soft scale $Q_s\equiv\sqrt{-t}\ll Q$.
The {\it exclusive} processes allow to extract the hadron's 
generalized parton distributions (GPDs)~\cite{Muller:1994ses, Ji:1996ek, Radyushkin:1997ki, Goeke:2001tz, Diehl:2003ny}, $F(x, \xi, t)$, 
which depend on the $t$ and hadron momentum skewness $\xi = [(p-p')\cdot n] / [(p+p')\cdot n]$ along a chosen light-cone direction $n$ 
in addition to longitudinal momentum fraction $x$ of the active parton~\cite{Qiu:2022pla}.
Unlike the TMDs, no perturbative showering dilutes the hadron structure information encoded in GPDs.  
The Fourier transform of the GPDs' dependence on the soft scale $t$ at $\xi\to 0$ limit reveals the images of partons inside a confined hadron
$f(x, \bm{b}_T)$ at transverse spatial positions $b_T$ in slices of different $x$~\cite{Burkardt:2000za, Burkardt:2002hr}. 
Also, the $x$-moments of GPDs $F_n(\xi, t) = \int_{-1}^1 dx x^{n-1} F(x, \xi, t)$ connect to various emergent properties of hadrons,
including their mass~\cite{Ji:1994av, Ji:1995sv, Lorce:2017xzd, Metz:2020vxd} and spin~\cite{Ji:1996ek} composition, 
internal pressure, and shear force~\cite{Polyakov:2002yz, Polyakov:2018zvc, Burkert:2018bqq}.

Owing to the QCD color confinement, no quarks and gluons can be detected in isolation. 
It is the QCD factorization that matches the probed hadron to its internal partonic structure, 
making it possible to extract the partonic information from experimental measurements with controllable approximations.  
Since multiple subprocesses with different TMDs or GPDs, distinguished by their spin and parity properties, 
can contribute to the same two-scale cross section, 
extracting the internal partonic structure information from experimental data is a challenging inverse problem. 
For two-scale {\it exclusive} observables, cross sections could also receive contributions from subprocesses independent of GPDs, 
making the extraction of GPDs even more difficult than TMDs~\cite{Ji:1996nm, Radyushkin:1997ki, Brodsky:1994kf, Frankfurt:1995jw, Berger:2001xd, Guidal:2002kt, Belitsky:2002tf, Belitsky:2003fj, Kumano:2009he, Kumano:2009ky, ElBeiyad:2010pji, Pedrak:2017cpp, Pedrak:2020mfm, Siddikov:2022bku, Siddikov:2023qbd, Siddikov:2024blb, Boussarie:2016qop, Duplancic:2018bum, Duplancic:2022ffo, Duplancic:2023kwe, Qiu:2022bpq, Qiu:2023mrm, Qiu:2024mny}.  
The fact that QCD factorization of GPDs is carried out at the amplitude level
makes the partonic momentum fraction $x$ a variable of virtual loop momentum, always integrated from $-1$ to $1$, 
while the $\xi$ and $t$ are directly observed hadronic variables.
This makes the extraction of $x$-dependence of GPDs a very difficult inversion problem~\cite{Bertone:2021yyz}. 

In the following, we will briefly review the recently developed framework for extracting GPDs 
from both existing and new two-scale {\it exclusive} processes in a uniform and coherent way, 
providing some ideas and opportunities to overcome the challenges of separating different GPDs and extracting their $x$-dependence.

\section{Single-diffractive hard exclusive processes for studying GPDs}
\label{sec:sdhep}
\begin{figure}[htbp]
	\centering
	\begin{tabular}{cc}
		\includegraphics[trim={-1em -4em -1em 0}, clip, scale=0.6]{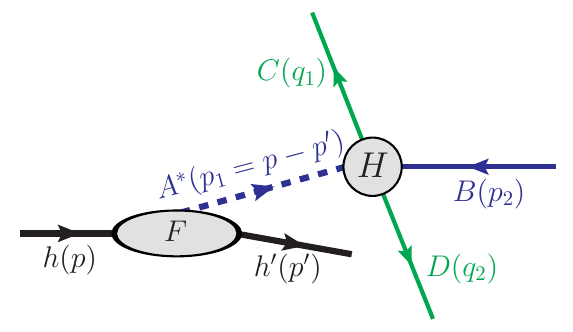} &
		\includegraphics[trim={-1em 0 -1em 0}, clip, scale=0.45]{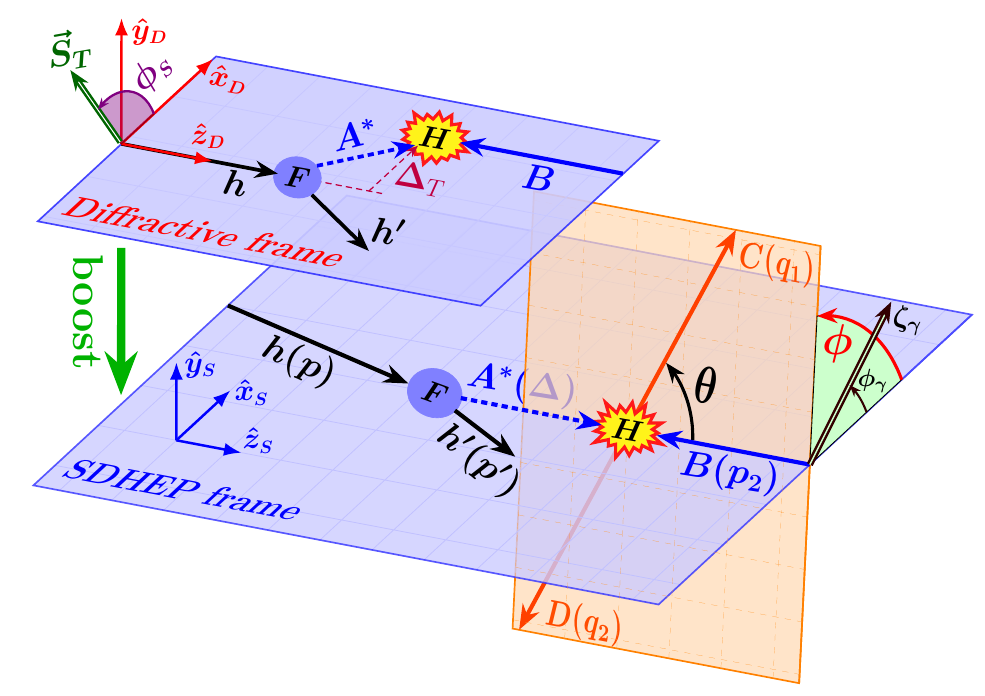} \\
		(a) & (b)
	\end{tabular}
	\caption{(a) Dynamic separation of the SDHEP amplitude into two stages. 
	    (b) The frame for analyzing SDHEPs in \eq{eq:sdhep}, where
		the linear polarization $\zeta_{\gamma}$ along $\phi_{\gamma}$ applies only to photoproduction processes.
		}
	\label{fig:sdhep}
\end{figure}

Because all physical processes that can be used to extract GPDs have a common feature of 
involving a diffraction of the colliding hadron $h$ to the observed hadron $h'$ and a hard interaction as a short-distance probe, 
the minimal kinematic configuration requires a $2 \to 3$ exclusive process,
\beq[eq:sdhep]
	h(p) + B(p_2) \to h'(p') + C(q_1) + D(q_2),
\eeq
with another colliding particle $B$ and produced particles $C$ and $D$, whose large balancing transverse momenta 
(with respect to the $h$-$B$ collision axis) defines the hard scale $Q\equiv q_T \sim q_{1T} \sim q_{2T} \gg \sqrt{-t}$.
We refer to such a $2 \to 3$ process as a single-diffractive hard exclusive process (SDHEP)~\cite{Qiu:2022pla}, 
which has a genetic two-stage feature,
\bse\label{eq:two-stage}\begin{align}
	&h(p) \to A^*(\Delta = p - p') + h'(p'), 	\label{eq:diffractive}\\
    &\hspace{9ex} \begin{tikzpicture}
        \node[inner sep=0pt] (arrow) at (0, 0) {
            \tikz{\draw[->, >={Stealth}, double, double distance=1pt, line width=1pt] (0, 0.28) to [out=-90, in=180] (0.8, 0);}
        };
    \end{tikzpicture} \hspace{1ex}
    A^*(\Delta) + B(p_2) \to C(q_1) + D(q_2),  \label{eq:hard 2to2}
\end{align}\ese
where the diffraction in (\ref{eq:diffractive}) and the exclusive $2\to 2$ hard collision in (\ref{eq:hard 2to2}) 
are connected by a virtual state $A^*$ of momentum $\Delta=p-p'$, as shown in Fig.~\ref{fig:sdhep}(a), 
whose invariant mass $\sqrt{-t}$ is much smaller than its energy $\sim {\cal O}(q_T)$.  
That is, the virtual state $A^*$ is much more long-lived than the time scale of the hard collision ($\sim 1/q_T$), 
which is necessary for the factorization of soft dynamics 
taking place at the scale $\sqrt{-t}$ from that of the hard collision that serves as a short-distance probe.

As shown in \fig{fig:sdhep}(b), we describe the diffraction subprocess (\ref{eq:diffractive}) using $(t, \xi, \phi_S)$ in the {\it diffractive frame},
where $h$ and $B$ collide along a collinear $\hat{z}_D$ axis.
The $\hat{x}_D$ axis is chosen in the diffractive plane along the transverse momentum $\bm{\Delta}_T$ of the $A^*$.  
It varies from event to event, and is compensated by 
the varying azimuthal angle $\phi_S$ of the hadron transverse spin $\bm{S}_T$, which is fixed in the lab frame. 
In this diffractive frame, the large components of the hadron momenta are selected by the lightlike vector $n = (1, 0, 0, -1) / \sqrt{2}$ along $-\hat{z}_D$,
which is needed to define the skewness $\xi$ independent of the details of the hard collision.

The $\xi$ and $t$ determine the c.m.~energy squared, $\hat{s} = (\Delta + p_2)^2$, of the $2\to2$ hard scattering in \eq{eq:hard 2to2}.
We refer to the c.m.~frame of the hard collision as the {\it SDHEP frame},
where the $\hat{z}_S$ axis is along the direction of $A^*$ and the $\hat{y}_S \propto \bm{p}' \times \bm{p}$ perpendicular to the diffraction plane.
This transformation from the diffractive frame to the SDHEP frame is characterized by a power suppressed transverse boost~\cite{Diehl:2003ny}
which leaves $n$ and thus the definition of $\xi$ invariant.
In this way, $\xi$ is both a variable of GPDs and a kinematic observable for SDHEPs.
Each event is then described by the five kinematic variables $(t, \xi, \phi_S, \theta, \phi)$,
where $\theta$ and $\phi$ are the polar and azimuthal angles of the observed particle $C$ in the SDHEP frame.

Notably, the SDHEP in \eq{eq:sdhep} covers all the known processes for extracting GPDs. 
For an electron beam, $B = e$, we have the electroproduction of a real photon [with $(CD) = (e\gamma)$] and a light meson [with $(CD) = (e M)$].
For a photon beam, $B = \gamma$, we have the photoproduction of dilepton [with $(CD) = (\ell^-\ell^+)$], diphoton [with $(CD) = (\gamma\gamma)$],
photon-meson pair [with $(CD) = (M \gamma)$], and meson-meson pair [with $(CD) = (M_1 M_2)$].
For a meson beam, $B = \pi$ or $K$, we have the mesoproduction of dilepton [with $(CD) = (\ell^-\ell^+)$] and diphoton [with $(CD) = (\gamma\gamma)$].
Besides these, one may easily conjecture new processes, such as the mesoproduction of 
photon-meson pair [with $(CD) = (M \gamma)$], and meson-meson pair [with $(CD) = (M_1 M_2)$], etc.

Dynamically, 
the scattering amplitude $\M$ can be formally written as a product or convolution of the amplitudes of the two subprocesses, 
summing over all possible $A^*$ states,
\beq[eq:channels]
    \M^{hB\to h'CD}_{\lambda_h \lambda_B} = \sum_{A^*} 
	F^{h \to h'A^*}_{\lambda_h}(t, \xi, \phi_S) \otimes
    H_{\lambda_A \lambda_B}^{A^*B \to CD}(\hat{s}, \theta, \phi),
\eeq
where the sum is over different numbers, $m$, and flavor contents of the exchanged particles in $A^*$,
$F^{h \to h'A^*}_{\lambda_h}$ describes the diffraction with $\lambda_h$ being the helicity of $h$ in the diffractive frame,
and $H_{\lambda_A \lambda_B}^{A^*B \to CD}$ stands for the hard $2\to2$ scattering amplitude 
with $\lambda_A$ and $\lambda_B$ the helicities of $A^*$ and $B$ in the SDHEP frame, respectively.
This two-stage description separates the physics taking place at different scales $\sqrt{-t}$ and $q_T$, 
with the size of contribution from $m$-th channel scaled as $(\sqrt{-t} / q_T)^{m-1}$ relative to the channel at $m = 1$.
For the leading channel, $m = 1$, $A^*$ can only be a virtual photon $\gamma^*$, 
which probes the electromagnetic form factors ($F_1$ and $F_2)$ of $h$.
For the subleading channel with $m = 2$, we have two-parton states, $A^* = [q\bar{q}]$ or $[gg]$, 
that form a loop to connect the diffraction and hard scattering subdiagrams in \fig{fig:sdhep}. 

At the leading power of $\sqrt{-t} / q_T$, the major contribution at $m = 2$ comes from the loop momentum integration 
when the two active partons propagate almost collinearly.
This picture is similar to the exclusive large-angle $2\to2$ scattering of hadrons~\cite{Lepage:1980fj, Brodsky:1989pv},
where the various soft gluon exchanges are canceled because each collinear pair of partons 
are so close to each other to behave effectively like a color-singlet state when approaching to the hard collision.
Generalizing to the single diffractive processes studied here, one can similarly show that any soft gluons 
connected to $B$, $C$, and/or $D$ are cancelled 
to the leading power of $\sqrt{-t} / q_T$~\cite{Qiu:2022pla}.
Although the diffractive kinematics contains the DGLAP region that partially pinches the soft gluon momentum in the Glauber region, 
such cancellation decouples the soft gluons from other collinear sectors,
resulting in the factorization of GPDs from the $2\to 2$ hard subdiagram, 
\beq[eq:factorization]
	\mathcal{M}^{[n = 2]}_{hB\to h'CD} = \sum_{i} \int_{-1}^1 dx \, F_{hh'}^{i}(x, \xi, t; \phi_S) \, G_{i B \to CD}(x, \xi; \hat{s}, \theta, \phi)
		+ \mathcal{O}\pp{\sqrt{-t} / q_T},
\eeq
where the $\sum_i$ runs over both parton flavors and GPD types.
The contributions from $m \geq 3$ channels are further suppressed by powers of $\sqrt{-t} / q_T$. 

Similarly to the factorization of exclusive large-angle $2\to2$ scattering in \refs{Lepage:1980fj, Brodsky:1989pv}, 
the GPD factorization of SDHEPs generally assumes 
the suppression of the endpoint region where one of the active partons from the meson(s) in $\{B, C, D\}$ becomes soft
and that the $2\to2$ subprocess in \eq{eq:hard 2to2} is dominated by one single hard scattering.
Independently of these assumptions, though, it is easy to see that double-diffractive hard exclusive processes are not factorizable 
due to the pinch of soft gluons in the Glauber region when the partons of both diffracted hadrons are in the DGLAP region.
That is, the generalization of SDHEPs to $2\to4$ processes and beyond
needs to stay in the single diffractive mode while the hard $2\to2$ subprocess in \eq{eq:hard 2to2} can be modified to $2\to 3$, etc.

\section{Azimuthal modulations and separation of different GPDs}
\label{sec:azimuth}
As discussed above, the leading channel with $A^* = \gamma^*$ at $m = 1$, 
generally referred to as Bethe-Heitler (BH) subprocess,
has no sensitivity to GPDs.
Their contribution needs to be separated from that of GPDs at the cross section level,
which is not trivial since both contributions interfere at the amplitude level.
Conventionally, the BH subprocess has been treated individually as a ``background'' of the GPD-sensitive subprocess.
The analyzing frame is designed for the latter and differs from process to process.
As a result, the choice of the light-cone vector $n$ to define the $\xi$ and thereby the definition of GPDs differ from process to process.
In addition, the BH subprocess complicates the azimuthal modulations needed to separate contributions from different GPDs.  
While the frame choice does not alter the underlying dynamics, it does affect the distribution of the observed variables. 
Since the BH and GPD subprocesses as well as different GPDs contribute to the same cross section,
it is important to choose a frame in which the azimuthal distribution directly reflects the quantum numbers of the exchanged state $A^*$ 
that defines the different GPDs.

In terms of the SDHEP framework, as described in \sec{sec:sdhep}, 
the BH and GPD subprocesses are just the first two leading channels of $A^*$,
so both are naturally described using the two-stage picture in the diffractive and SDHEP frames, respectively.
In both frames, the relevant initial states for the azimuthal angle $\phi_S$ or $\phi$ are put along the $z$ axis,
so the dependence of the amplitude on $\phi_S$ or $\phi$ is simply determined by the initial-state helicity,
\begin{align}
	F^{h \to h'A^*}_{\lambda_h}(t, \xi, \phi_S) 
		&= e^{-i \lambda_h \phi_S} F^{h \to h'A^*}_{\lambda_h}(t, \xi, 0), \nn\\
    H_{\lambda_A \lambda_B}^{A^*B \to CD}(\hat{s}, \theta, \phi)
    		&= e^{i\pp{\lambda_{A} - \lambda_B} \phi} H_{\lambda_A \lambda_B}^{A^*B \to CD}(\hat{s}, \theta, 0).
\end{align}
When computing physical observables, one needs to square \eq{eq:channels} and trace over the spin density matrices for the incoming particles, i.e., 
$ \sum\nolimits_{\{\lambda\}} 
	\rho^{(h)}_{\lambda_h\lambda'_h} \rho^{(B)}_{\lambda_B\lambda'_B} \M_{\lambda_h \lambda_B} \M_{\lambda'_h \lambda'_B}^*$, 
which causes different $h$ and $B$ helicities as well as different $A^*$ channels to interfere, 
giving rise to a variety of azimuthal modulations in $\phi_S$ and $\phi$. 
For instance, for a nucleon target with transverse spin $S_T$, the interference of $\lambda_h = \pm 1/2$ leads to $\cos\phi_S$ and $\sin\phi_S$ modulations,
while the interference between two $(A^*, B)$ channels of helicities $(\lambda_A, \lambda_B)$ and $(\lambda_A^{\prime}, \lambda'_B)$ would lead to the azimuthal modulations 
$\cos[ (\Delta\lambda_A - \Delta\lambda_B) \phi ]$ and $\sin[ (\Delta\lambda_A - \Delta\lambda_B) \phi ]$,
with $(\Delta\lambda_A, \Delta\lambda_B) \equiv (\lambda_A - \lambda_A', \lambda_B - \lambda'_B)$. 

It is an important observation that $\phi$ modulations arise from the interference of different $A^*$ channels.
Since the BH has $\lambda_A^{\gamma} = \pm 1$ (at leading power)
while the unpolarized and polarized GPDs, $F$ and $\wt{F}$, respectively, both have $\lambda_A^{[q\bar{q}]} = \lambda_A^{[gg]} = 0$,
their interference yields $\cos\phi$ and/or $\sin\phi$ modulations as a clear signature of the GPDs.
The transversity GPDs $F_T$ give $\lambda_A^{[q\bar{q}]_T} = \pm 1$ or $\lambda_A^{[gg]_T} = \pm 2$,
whose interference with the $\lambda_A^{\gamma} = \pm 1$ can lead to more $\phi$ modulations,
similarly for high-twist GPDs. 
On the other hand, in the decomposition of the GPD $F$ into the $H$ and $E$ GPDs (similarly for $\wt{F}$ and transversity GPDs),
the $H$ and $E$ control different helicity configurations for the hadron diffraction, so they give different contributions to the $\cos\phi_S$ and $\sin\phi_S$ modulations
which can help separate them from each other~\cite{Qiu:2024pmh}.

As a specific example, we present the azimuthal modulation in the real photon electroproduction process (or, BH+DVCS)
within the next-to-leading-power accuracy (i.e., with $m \leq 2$),
\begin{align}
	\frac{d\sigma^{\gamma}_e}{dt \, d\xi \, d\phi_S \, d\cos\theta \, d\phi}
	&= \frac{1}{(2\pi)^2} \frac{d\sigma^{\gamma, {\rm unp.}}_e}{dt \, d\xi \, d\cos\theta} \,
    		\bigg[ 1 + P_e P_N A^{\rm LP}_{LL} + P_e S_T A^{\rm LP}_{TL}  \cos\phi_S \nn\\
		&\hspace{6em} 
			+ \pp{ A_{UU}^{\rm NLP} + P_e P_N A_{LL}^{\rm NLP} } \cos\phi 
			+ \pp{ P_e A_{UL}^{\rm NLP} + P_N A_{LU}^{\rm NLP} }\sin\phi		\nn\\
		&\hspace{6em} 
			+ S_T \pp{ A_{TU, 1}^{\rm NLP} \cos\phi_S \sin\phi + A_{TU, 2}^{\rm NLP} \sin\phi_S \cos\phi }		\nn\\
		&\hspace{6em} 
			+ P_e S_T \pp{ A_{TL, 1}^{\rm NLP} \cos\phi_S \cos\phi + A_{TL, 2}^{\rm NLP} \sin\phi_S \sin\phi}
		\bigg]
\label{eq:dvcs-xsec}
\end{align}
where $P_e$ and $P_N$ are the net helicities of the electron and nucleon beams, respectively.
The forms of the unpolarized differential cross section $d\sigma^{\gamma, {\rm unp.}}_e$ 
and azimuthal asymmetry coefficients $A$'s can be found in \refcite{Qiu:2024pmh}.
They depend in a simple way on the GPDs and can be extracted by azimuthal projections from data.

A similar result for the dilepton photoproduction process has also been worked out in \refcite{Qiu:2024pmh}.
For other processes, the BH channels are not as important so the azimuthal modulations arise from the interference of different GPDs 
or the $B$ helicities in the case of photoproduction processes.
The SDHEP framework gives not only a simple and clear azimuthal formulation, 
but also a consistent choice of the light-cone vector $n$ and GPD definitions,
and therefore the GPDs extracted from various processes can be directly compared.

\section{Extracting the $x$-dependence of GPDs}
\label{sec:x}

GPDs enter physical observables with the range of their $x$-integration from $-1$ to $1$.
For example, the asymmetries in \eq{eq:dvcs-xsec} depend only on such ``moment'' integrals of GPDs,
\beq[eq:gpd-moments]
    \big\{ \H, \, \E, \, \Ht, \, \Et \big\}(\xi, t) 
    = \sum_q e_q^2 \int_{-1}^1 dx \, \frac{\big\{ H^{q, +}, \, E^{q, +}, \, \wt{H}^{q, +}, \, \wt{E}^{q, +} \big\}(x, \xi, t)}{x - \xi + i \epsilon},
\eeq
where the `$+$' superscripts refer to charge-conjugation-even GPD combinations, 
\beq
    F^{q, +}(x, \xi, t)
    = F^{q}(x, \xi, t) \mp F^{q}(-x, \xi, t),
\eeq
with $\mp$ for $F = H$ or $E$ and $F = \wt{H}$ or $\wt{E}$, respectively.
While the proper choice of the SDHEP frame makes the azimuthal modulations simple, 
the extraction of GPDs is only up to such integrals that result from the dynamics of the particular process.
Since it is simple to construct analytic functions $S(x, \xi, t)$, called shadow GPDs~\cite{Bertone:2021yyz}, 
that satisfy all known features of GPDs, such as the polynomality, trivial forward limit, 
and giving zero contribution to the integral in \eq{eq:gpd-moments},
inverting an $x$-dependent GPD solution from \eq{eq:gpd-moments} is an ill-defined and singular problem.

This singularity is due to the leading-order scaling property of such GPD moments.
It is similar to the scaling phenomenon in inclusive processes and is a result of the fact that 
the internal propagators in the hard part are directly connected to two external lightlike momenta on either end~\cite{Qiu:2022pla}.
External observables like $\theta$ simply do not enter the hard kernel in the GPD convolution in \eq{eq:factorization}.
To have enhanced sensitivity to the GPD $x$-dependence, one thus needs to break the scaling at leading order.
This, unfortunately, does not happen to most of the known $2\to3$ SDHEPs listed in \sec{sec:sdhep}.
Two known $2\to3$ SDHEPs with non-scaling GPD integrals are the 
diphoton mesoproduction in nucleon-pion scattering~\cite{Qiu:2022bpq, Qiu:2024mny},
$N(p) + \pi(p_2) \to N'(p') + \gamma(q_1) + \gamma(q_2)$,
and  
photon-pion pair photoproduction in nucleon-photon collision~\cite{Duplancic:2018bum, Duplancic:2022ffo, Duplancic:2023kwe, Qiu:2022pla, Qiu:2023mrm},
$N(p) + \gamma(p_2) \to N'(p') + \pi(q_1) + \gamma(q_2)$.

In the hard scattering diagrams of the diphoton mesoproduction, the two photons can be radiated from different quark lines.
When $q_T\gg \sqrt{-t}$, the two quark lines must be connected by a hard gluon that transmits the $q_T$ flow.
This generates a new GPD integral besides \eq{eq:gpd-moments},
\begin{align}
	I(\xi, t; z, \theta)
	= \int_{-1}^{1} dx \frac{F^+(x, \xi, t)}{x - x_p(\xi, z, \theta) + i \epsilon \, \sgn{\cos^2(\theta/2) - z} },
\label{eq:diphoton-special-int}
\end{align}
with a new pole $x_p$ that depends on $\theta$ (or equivalently, $q_T$),
\beq[eq:x-pole]
	x_p(\theta)
		= \xi \bb{ \frac{1 - z + \tan^2(\theta / 2) \, z}{1 - z - \tan^2(\theta / 2) \, z} }.
\eeq
Varying $\theta$ shifts this pole around in the DGLAP region of the GPDs, 
causing an entanglement between the GPD $x$ dependence and the observable $\theta$ (or $q_T$) distribution.
The second process is related to the first one by a kinematic crossing, so contains a similar integral to \eq{eq:diphoton-special-int}, 
but with the pole moving in the ERBL region, so gives complementary sensitivity.
	
\begin{figure}[htbp]
	\centering
	\begin{tabular}{cc}
		\includegraphics[scale=0.46]{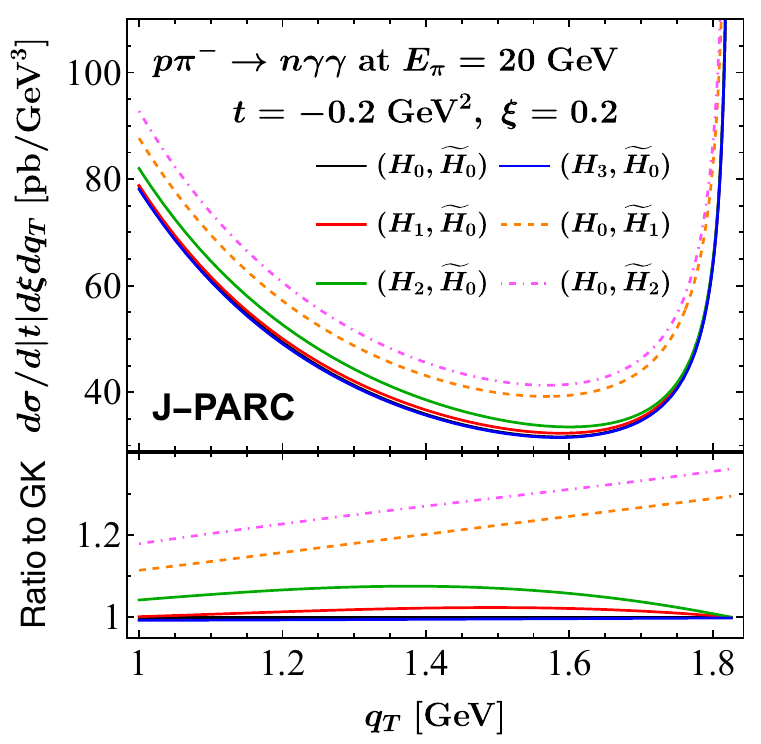} &
		\includegraphics[scale=0.52]{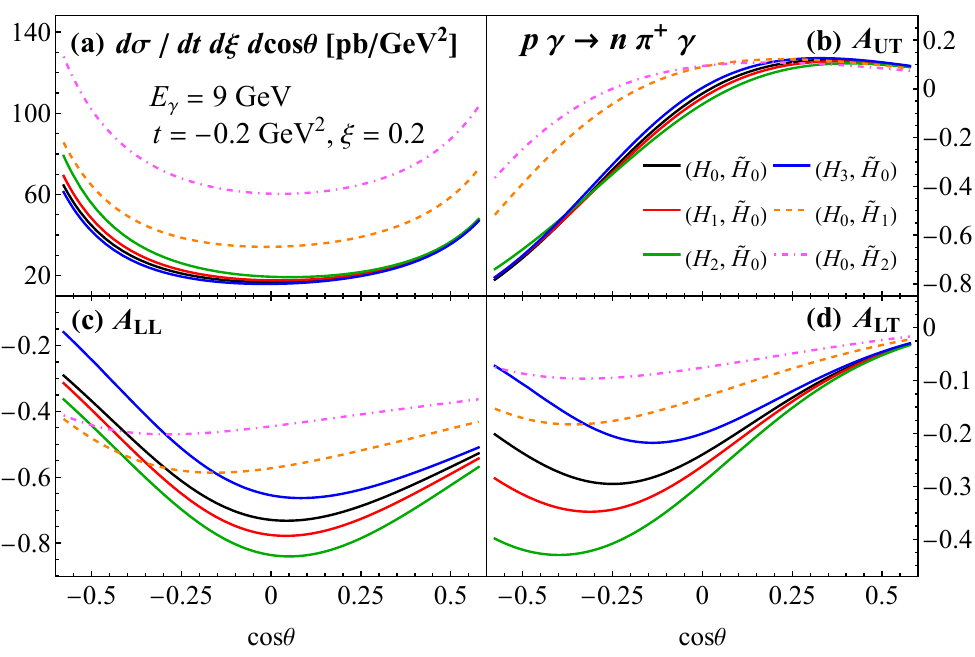} \\
		(a) & (b)
	\end{tabular}
	\caption{(a) The $q_T$ distributions for the diphoton cross section, and 
		(b) the $\cos\theta$ distributions for the photoproduction cross section and polarization asymmetries,
		evaluated using different GPD models.
	}
	\label{fig:qt-dist}
\end{figure}

Both processes can distinguish shadow GPDs from real ones. To demonstrate this, 
we construct some GPD models by starting with the Goloskokov-Kroll model~\cite{Goloskokov:2005sd, Goloskokov:2007nt, Goloskokov:2009ia, Kroll:2012sm} 
$(H_0, \wt{H}_0)$.
Adding analytically constructed shadow GPDs leads to a set of model GPDs
$(H_0, H_1, H_2, H_3)$ and $(\wt{H}_0, \wt{H}_1, \wt{H}_2)$, which give effectively the same contribution to the DVCS-type processes.  
We calculate the differential observables for both processes using these model GPDs.
For the diphoton mesoproduction, since pion is spin zero, one can only measure an unpolarized cross section (unless the proton carries a transverse spin).
This is shown in \fig{fig:qt-dist}(a) for the J-PARC kinematics.
For the photon-pion photoproduction, one can have additional polarization asymmetries when the photon is circularly or linearly polarized, 
giving more independent constraints on the GPDs, as shown in \fig{fig:qt-dist}(b) for the JLab Hall-D kinematics.
Clearly, in both cases, those different GPD models can be separated by measuring both the magnitude and shapes of the different observables.

Beyond $2\to3$ SDHEPs, one can readily have more processes at $2\to4$ level with non-scaling hard kernels.
Some simple examples are 
(1) dilepton electroproduction process (or double DVCS)~\cite{Guidal:2002kt, Belitsky:2002tf, Belitsky:2003fj, Deja:2023ahc}, $p + e \to p + e + \ell^+ + \ell^-$,
(2) diphoton electroproduction~\cite{Pedrak:2020mfm}, $p + e \to p + e + \gamma+ \gamma$, 
and 
(3) photon-dilepton pair photoproduction, $p + \gamma \to p + \gamma + \ell^+ + \ell^-$.
Most of such processes remain un-studied. 
They are generally more kinematically involved and more difficult to calculate or measure experimentally.

\section{Summary}
\label{sec:summary}
In terms of the SDHEP framework, we briefly reviewed the strategy to overcome the challenges for separating different GPDs and the extraction of their $x$-dependence. 
We have shown that using azimuthal modulations can help disentangle BH and GPD subprocesses as well as GPDs of different spin and parity structures.
For the inversion problem of extracting the $x$-dependence of GPDs, it is important to study processes that yield non-scaling hard parts in the GPD integrals.

The consistent and process-independent definition of GPDs in the SDHEP framework
paves the way towards a global analysis of GPDs from experimental data.
Similarly to that of parton distribution functions and TMDs, one also needs to combine multiple processes in the global analysis 
to disentangle different parton flavors and probe different charge conjugation GPD combinations.
Differently from parton distribution functions, though, GPDs also depend on two extra hadronic variables, $\xi$ and $t$.
It is therefore important to use experiments across a wide range of kinematics to extend the $\xi$ and $t$ coverage.

Because of the exclusive nature, while GPDs are defined as twist-2 distributions at the amplitude level, 
their appearance at the cross section level is equivalent to high-twist contributions.
As a result, the associated event rate drops quickly as one goes to higher energies, making a precision measurement very challenging.
On the other hand, due to the factorizability condition, $q_T \gg \sqrt{-t}$, a higher-energy experiment is needed to extend the measurements to the larger-$t$ region, 
enabling the reliable Fourier transform to derive the spatial distribution of partons inside a confined hadron.  
Overcoming this challenge is essential for a practical realization of the tomographic program from the reliable extraction of GPDs.

\vspace{3mm}
This work is supported in part by the U.S. Department of Energy (DOE) Contract No.~DE-AC05-06OR23177, 
under which Jefferson Science Associates, LLC operates Jefferson Lab.

\bibliographystyle{apsrev}
\bibliography{reference}

\end{document}